\documentclass{pasj00}
\draft

\begin{document}
\SetRunningHead{Nakatake et al.}{$T_{\rm sys}$ measurement by 2-bit A/D}
\Received{2008/07/19}
\Accepted{2010/08/24}

\title{A New System Noise Measurement Method Using a 2-bit Analog-To-Digital Converter}

\author{Aki \textsc{Nakatake}, Seiji \textsc{Kameno}, and Koji \textsc{Takeda} %
  \thanks{Send offprint requests for kameno@sci.kagoshima-u.ac.jp}}
\affil{Department of Physics, Faculty of Science, Kagoshima University, 1-21-35, Korimoto, Kagoshima, 890-0065}
\email{nakatake@astro.sci.kagoshima-u.ac.jp, kameno@sci.kagoshima-u.ac.jp, takeda@astro.sci.kagoshima-u.ac.jp}

%


\maketitle

\begin{abstract}
We propose a new method to measure the system noise temperature, $T_{\rm sys}$, using a 2-bit analog-to-digital converter (ADC).
The statistics of the digitized signal in a four-level quantization brings us information about the bias voltage and the variance, which reflects the power of the input signal.
Comparison of the variances in {\it hot} and {\it sky} circumstances yields $T_{\rm sys}$ without a power meter.
We performed test experiments using the Kagoshima 6-m radio telescope and a 2-bit ADC to verify this method.
Linearity in the power--variance relation was better than 99\% within the dynamic range of 10 dB.
Digitally measured $T_{\rm sys}$ coincided with that of conventional measurement with a power meter in 1.8-\% difference or less for elevations of $10^{\circ} - 88^{\circ}$.
No significant impact was found by the bias voltages within the range between $-3.7$ and $+12.8$\% with respect to the threshold voltage.
The proposed method is available for existing interferometers that have a multi-level ADC, and release us from troubles caused by power meters.
\end{abstract}

\section{Introduction}
System noise temperature ($T_{\rm sys}$) is an important parameter for amplitude calibration in radio astronomy observations.
This value represents the total equivalent system noise power, measured in the unit of Kelvin, including atmospheric radiation and attenuation, antenna noise, spill over, amplifier noise, loss in transmission, etc.
For most radio sources and receivers, the received signal is dominated by the system noise power.
Calibration for $T_{\rm sys}$ is necessary to measure the flux density or the antenna temperature of radio sources from the received signal.

A conventional $T_{\rm sys}$ measurement has been carried out using an analog power meter.
One can estimate $T_{\rm sys}$ by comparing the output signal powers between two or more circumstances with different input signals whose power is known.
When the telescope points toward off-source blank sky, the output power, $W_{\rm sky}$, is given as $W_{\rm sky} = k_{\rm B} G B T_{\rm sys}$, where $k_{\rm B}$ is the Boltzmann constant, $G$ is the gain of the receiving system, and $B$ is the bandwidth.
In the case that the receiving system is covered with an absorber at the temperature $T_{\rm hot}$, i.e. {\it ``hot load''}, the output power, $W_{\rm hot}$, will be $W_{\rm hot} = k_{\rm B} G B (T_{\rm sys} + T_{\rm hot})$.
Here $T_{\rm hot}$ is the absorber temperature that is approximately equal to the room temperature of $~\sim 293$ K.
Using these two measurements we can estimate $T_{\rm sys}$ as
\begin{eqnarray}
T_{\rm sys} = \frac{W_{\rm sky}}{W_{\rm hot} - W_{\rm sky}}T_{\rm hot}. \label{eqn:tsys_power}
\end{eqnarray}

Precision, stability, and dynamic range are crucially required for the power meter to attain accurate $T_{\rm sys}$.
The accuracy in $T_{\rm sys}$ depends on the precision of the power meter.
Since $W_{\rm hot}$ is usually several times greater than $W_{\rm sky}$, the power meter must keep linearity in a power range as wide as several dB.
A small non-linear characteristic of the power meter can result in systematic error in $T_{\rm sys}$.

It is troublesome to maintain high precision and linearity in an analog sensor eternally, especially for radio telescopes in inaccessible environments such as space orbits, polar regions, and high-altitude plateaus.
If a power meter device can be eliminated by an alternative method to measure $T_{\rm sys}$, we can reduce costs, troubles, maintenance frequency, etc.

We propose a simple method to measure $T_{\rm sys}$ using an analog-to-digital converter (ADC), instead of a power meter.
Most of modern interferometers and radio telescopes are equipped with a high-speed ADC in each antenna element to acquire received signal in digital form that is processed in correlators and spectrometers.
Utilization of digitized signal requires no additional costs.
Since the digitized signal responds to voltage of the analog signal, its statistics represents the mean and variance (i.e. power) of the input signal.
If the statistics reflects sufficiently accurate and reliable power in a sufficiently wide power range, ADCs can be the alternative.

ADCs for radio interferometers sample the input analog voltage and quantize into a few---two, three, four, or more---levels.
Distribution of samples in these levels holds statistical information such as mean and variance of the input signal.
Coarse quantization can affect the performance in power measurements; two-level quantization is obviously degenerate in amplitude of the analog voltage and is useless in $T_{\rm sys}$ measurements. 
Four-level (2-bit) quantization holds some information about amplitude.
This quantization is commonly applied in radio interferometers such as  ALMA \citep{2002PASJ...54L..59O} and CARMA \citep{2006ASPC..351..157R}, and VLBI such as VERA \citep{2003ASPC..306..367K}, JVN \citep{2006evn..confE..71D}, VSOP \citep{1998Sci...281.1825H}, and VLBA \citep{1988IAUS..129..461R}.

In this paper we propose a method of measuring $T_{\rm sys}$ using a four-level (2-bit) ADC.
Realistic performance of the method, tested with the Kagoshima 6-m radio telescope \citep{1994ASPC...59...64O}, is also reported.

\section{Method}\label{sec:method}
\begin{figure}
  \begin{center}
    \FigureFile(80mm,40mm){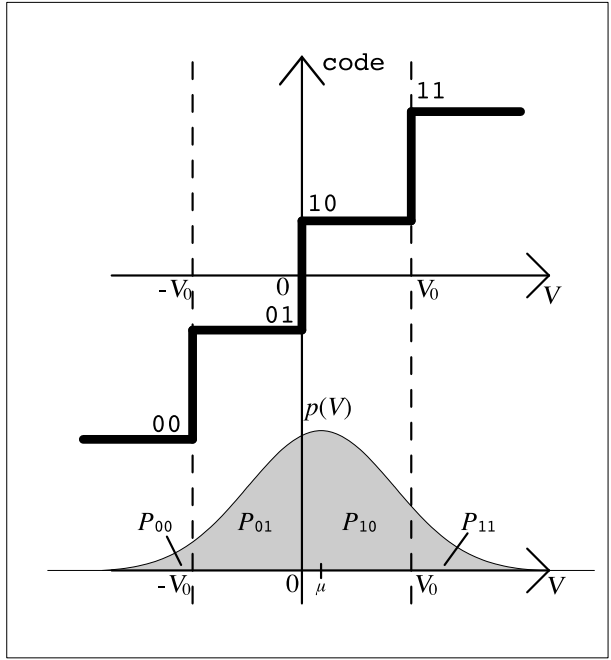}
  \end{center}
  \caption{Four-level encoding and probability density distribution. Horizontal axis indicates the input (analog) voltage. {\it Top} diagram shows the relation between the input voltage and the output codes {\tt 00, 01, 10,} and {\tt 11}, with the thresholds of $-V_0$, $0$, and $V_0$. {\it Bottom} diagram shows the probability density function, assuming the normal distribution with the mean and standard deviation of $\mu$ and $\sigma$. Probabilities of the output codes corresponds to the filled area in each level.}
  \label{fig:quantization_diagram}
\end{figure}

Thresholds in four-level quantization of ADCs are set at the voltages of $\pm V_0$ and $0$, and the input analog voltages are encoded into {\tt 00, 01, 10,} and {\tt 11} as shown in
figure \ref{fig:quantization_diagram} \citep{2001isra.book.....T}.
Distribution of samples in these codes reflects power and  bias of input signals.

We assume that the input signal is white noise whose probability density function of the voltage, $p(V)$, follows the normal distribution with the mean (bias) voltage of $\mu$ and the standard deviation of $\sigma$ as
\begin{eqnarray}
p(V) = \frac{1}{\sigma \sqrt{2\pi}} \exp \left[-\left( \frac{V-\mu}{2\sigma^2}\right)^2 \right].
\end{eqnarray}
This assumption is justified for white noise that has a flat spectrum across whole bandwidth \citep{2002papoulis}.
The probability in each digitized code will be
\begin{eqnarray}
P_{\tt 00} &=& \frac{1}{{\sigma\sqrt{2\pi}}}\int\limits_{-\infty}^{-V_0}\exp \left( -\frac{(V-\mu)^2}{2\sigma^2}\right) dV =\frac{1}{2}-\frac{1}{2}{\mathrm{erf}} \left( \frac{a_0+a_1}{\sqrt{2}}\right), \nonumber \\
P_{\tt 01} &=& \frac{1}{{\sigma\sqrt{2\pi}}}\int\limits_{-V_0}^{0}\exp \left(-\frac{(V-\mu)^2}{2\sigma^2} \right) dV = \frac{1}{2}{\mathrm{erf}}\left(\frac{a_0+a_1}{\sqrt{2}}\right)-\frac{1}{2}{\mathrm{erf}} \left( \frac{a_1}{\sqrt{2}}\right), \nonumber \\
P_{\tt 10} &=& \frac{1}{{\sigma\sqrt{2\pi}}}\int\limits_{0}^{V_0}\exp \left(-\frac{(V-\mu)^2}{2\sigma^2} \right) dV =\frac{1}{2}{\mathrm{erf}}\left(\frac{a_0-a_1}{\sqrt{2}}\right)+\frac{1}{2}{\mathrm{erf}} \left( \frac{a_1}{\sqrt{2}}\right), \nonumber \\
P_{\tt 11} &=& \frac{1}{{\sigma\sqrt{2\pi}}}\int\limits_{V_0}^{\infty }\exp \left(-\frac{(V-\mu)^2}{2\sigma^2} \right) dV =\frac{1}{2}-\frac{1}{2}{\mathrm{erf}} \left( \frac{a_0-a_1}{\sqrt{2}}\right), \label{eqn:probability_equations}
\end{eqnarray}
where \[ a_0=\frac{V_0}{\sigma}, \ a_1=\frac{\mu}{\sigma}, \ {\rm and } \ {\mathrm{erf}(x)}=\frac{2}{\sqrt{\pi}}\int\limits_{0}^{x}{\mathrm{exp}}(-t^2)dt. \]
These probabilities are constrained as  $P_{\tt 00} + P_{\tt 01} + P_{\tt 10} + P_{\tt 11}=1$.
The standard deviation, $\sigma$, and the bias voltage, $\mu$, can be estimated by using the three independent observable values of $P_{\tt 00}$, $P_{\tt 01}$, and $P_{\tt 10}$.
These probabilities are estimated by statistics of quantized signal with an ADC as $P_{\tt ml} = n_{\tt ml}/n$, when large enough number of samples are accumulated in each quantization level.
Here, $n_{\tt ml}$ is the number of samples in the quantization level of {\tt ml} ({\tt m} and {\tt l} stand for the most and least significant bits, respectively) and $n$ is the total number of samples.
Non-linear least squares analysis works to obtain the parameters of $a_0$ and $a_1$ from the ADC statistics of $P_{\tt 00}$, $P_{\tt 01}$ and $P_{\tt 10}$.

The power of the input signal is proportional to the variance, $\sigma^2$, as $\sigma^2 = \alpha W$, where $\alpha$ is a proportional constant.
Thus, equation \ref{eqn:tsys_power} will be expanded as
\begin{equation}
\hat{T}_{\rm sys} = \frac{\sigma_{\rm sky}^{2}}{\sigma_{\rm hot}^2 -\sigma_{\rm sky}^2 } T_{\rm hot} = \frac{(1/a_{0})_{\rm sky}^{2}}{ (1/a_{0})_{\rm hot}^2 -(1/a_{0})_{\rm sky}^2 } T_{\rm hot}. \label{eqn:tsys_adc}
\end{equation}
Here, we denote $\hat{T}_{\mathrm{hot}}$ as a system noise temperature that is measured by the ADC statistics.
Equation \ref{eqn:tsys_adc} indicated that $\hat{T}_{\mathrm{hot}}$ is calculated by $a_{0}$, that can be derived from ADC statistics, in {\it hot} and {\it sky} circumstances.

\section{Tests and Results}
We examined the new $T_{\rm sys}$ measurement by comparing with the conventional method using a power meter in four aspects---linearity, precision, dynamic range and robustness against input bias voltage. 
Linearity is estimated by relationship between the variance, $\sigma^2$ derived from the ADC statistics and the power of input signal measured by a power meter.
The proportional coefficient, $\alpha$, should be constant within a wide range of input power in an ideal linear system.
Dynamic range is evaluated by the width of the range where the system keeps proper linearity.
Precision of digital $\hat{T}_{\rm sys}$ is evaluated by departures from the conventional $T_{\rm{sys}}$ value measured simultaneously in the same configuration. 
Robustness against input voltage bias is tested by variation of $\sigma^2$ obtained from bit distribution of ADC, for various input bias voltage in ADC.

\begin{figure}
  \begin{center}
    \FigureFile(80mm,60mm){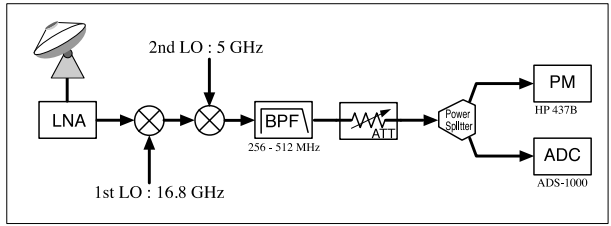}
  \end{center}
  \caption{Test configuration. The Kagoshima 6-m telescope pointed various elevations of the blank sky. Received signal was amplified in the low-noise amplifier (LNA) and was downconverted with the 1st and 2nd local oscillators (LOs) at 16.8 and 5 GHz to an IF signal. The IF signal was filtered in the bandpass filter (BPF) and bifurcated by the power splitter. It was quantized by the analog-to-digital converter (ADC). The IF power was measured by the power meter (PM) in analog form at the same time. The hot load can cover a feed horn inside the telescope. The IF power was adjusted at a desired level by the attenuator (ATT).}
  \label{fig:TestConfig}
\end{figure}

We used the Kagoshima 6-m radio telescope for these test experiments.
The configuration is shown in figure \ref{fig:TestConfig}.
The antenna pointed to blank sky at various elevations to generate system noise.
The feed horn inside the antenna was covered with a hot load when we measured $W_{\rm hot}$.
Received signal at 22.2 GHz was downconverted to intermediate frequency (IF) with the 1st and 2nd local oscillators at 16.8 and 5 GHz, respectively.
A step attenuator was used to adjust the IF power at a desired level.
The IF signal at 256 -- 512 MHz was bifurcated into two streams.
One signal was digitized by the analog-to-digital converter (ADC), ADS--1000 \citep{2001ExA....11...57N}, at a sampling frequency of 512 Msps with four-level quantization.
The other was used to measure the power in analog form with the power sensor, HP 8481A attached to the power meter, HP 437B.

Four experiments were carried out to test the aspects mentioned above. 

\subsection{Distribution of codes}
\begin{figure}[h!]
\begin{center}
\FigureFile(80mm,50mm){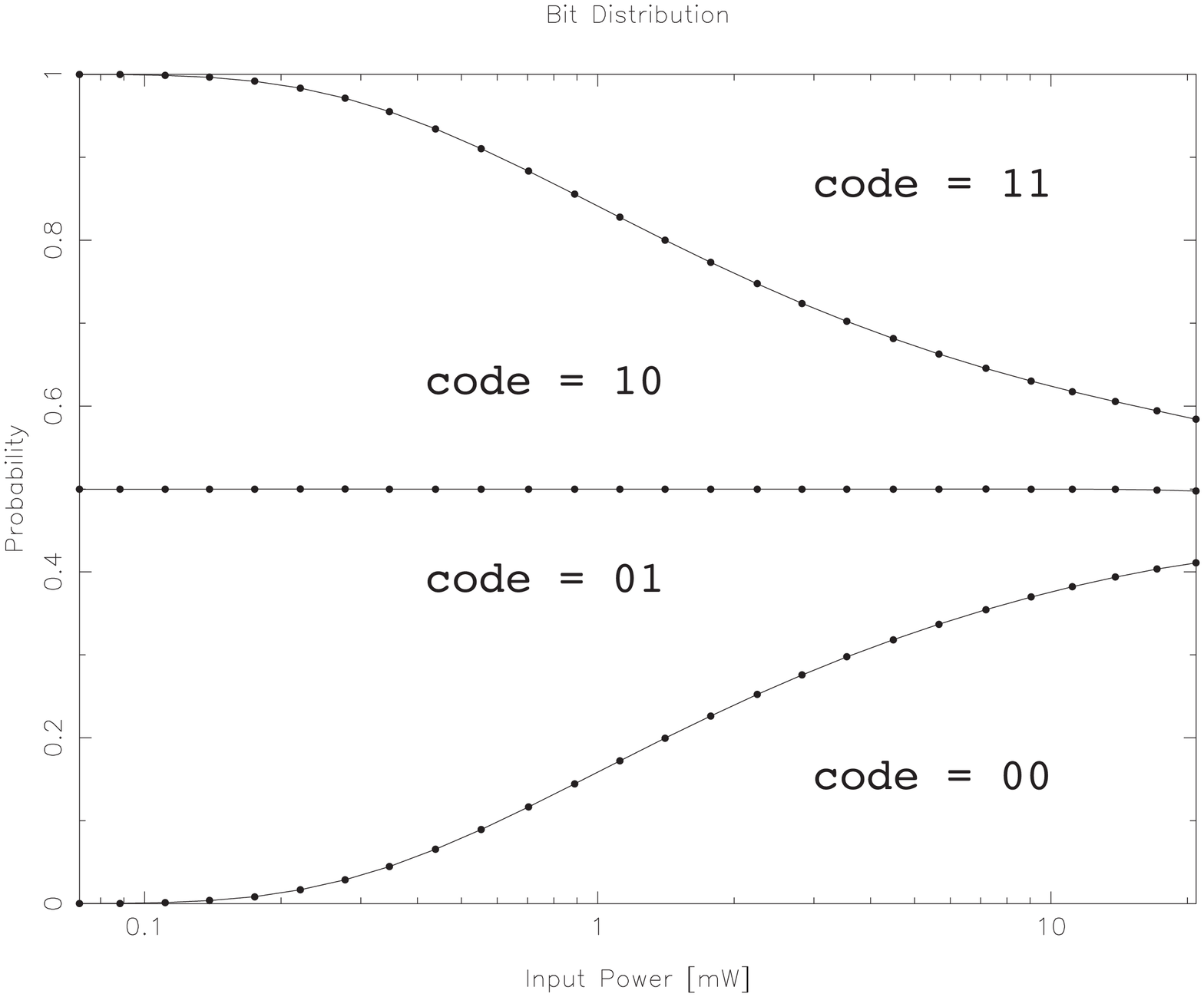}
\FigureFile(80mm,50mm){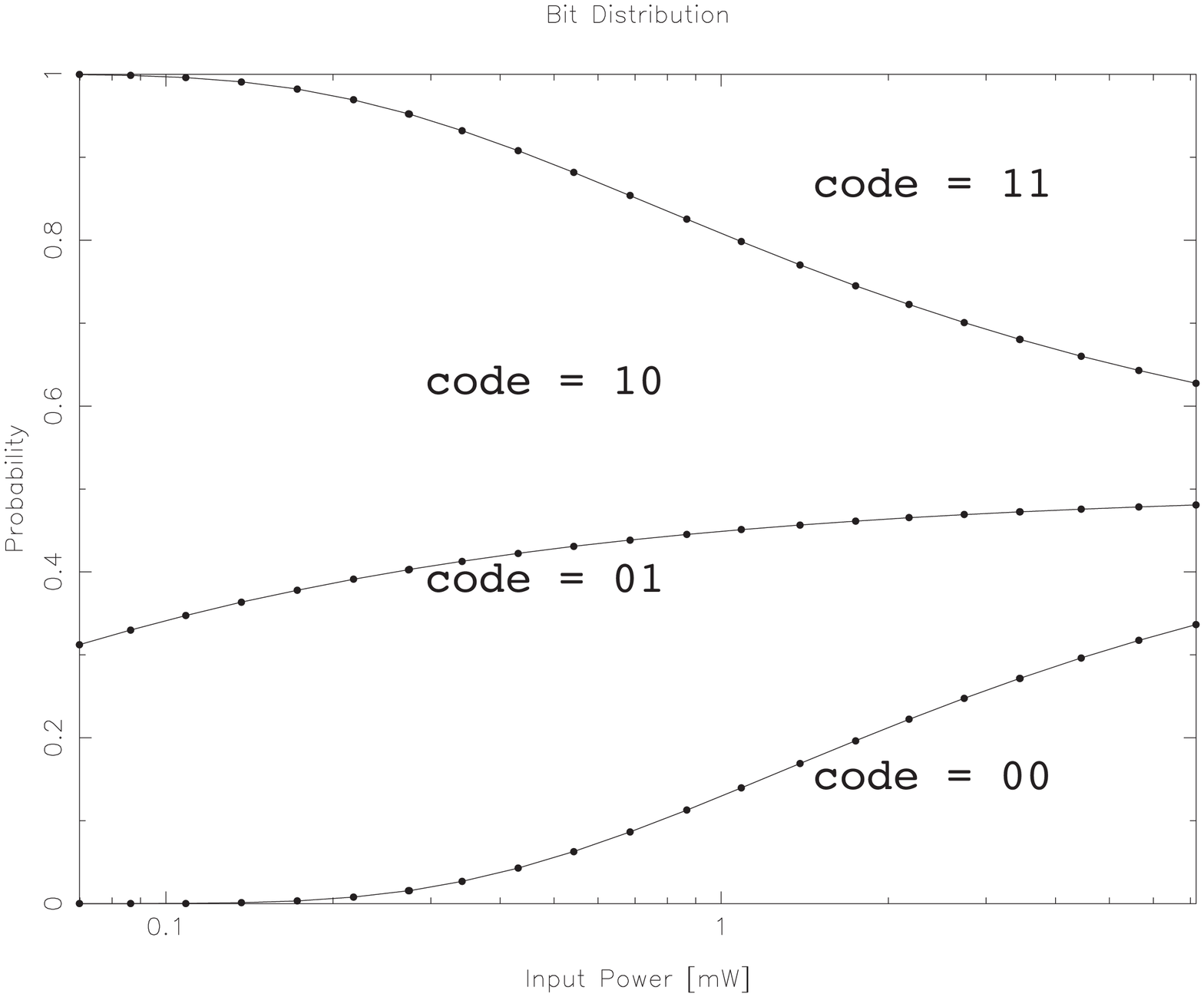}
\caption{($a$) Distribution of acquired samples coded as {\tt 00}, {\tt 01}, {\tt 10}, and {\tt 11} without bias offset.
Shares in the outer codes of {\tt 00} and {\tt 11} were 0.00949 \% and 0.00975 \%, respectively, at the lowest input power of 0.07 mW and increased at higher input power.

($b$) Distribution with a bias of $\mu/V_0 = 0.13$.
The distribution became significantly asymmetric across the codes.
Shares in {\tt 00} and {tt 11} were 0.00097 \% (i.e. 4963 samples in 1 sec) and 0.04864 \%, respectively.
}
\label{fig:bitdist}
\end{center}
\end{figure}

We examined statistics of samples coded in {\tt 00}, {\tt 01}, {\tt 10}, and {\tt 11} in various input power and bias offset.
The total number of samples was $512 \times 10^6$ in 1-sec accumulation.
Figure \ref{fig:bitdist} shows distributions of acquired samples, as a function of input power, without bias and with the maximum bias offset of $\mu/V_0 = 0.13$.

The distribution was symmetric without bias and became significantly asymmetric with the bias.
Percentages of upper levels of {\tt 00} and {\tt 11} were 0.00949 \% (48570 samples) and 0.00975 \% (49925 samples) without a bias and 0.00097 \% (4963 samples) and 0.04864 \% (249027 samples) with the maximum bias offset.
These shares increased at higher input power, as expected in equations \ref{eqn:probability_equations}.

\subsection{Power and digital variance}
To verify the linearity between the input analog power and the digital variance, $\sigma^2$, we compared them in various input power range by using a step attenuator between the bandpass filter (BPF) and the power splitter to adjust input signal power.
The input signal power was varied in the range from 0.03 mW ($-15$ dBm) through 7.9 mW ($+9$ dBm) with a 1-dB interval for the zero bias offset case.

For the case with a bias offset of $\mu / V_0 = 0.13$, the maximum input power was capped at 2.8 mW ($4.7$ dB) to avoid saturation.
We realized that this limitation was far from saturation and was not necessary after the measurement.
However, the linearity in the biased case was grasped with the capped power range.

Figure \ref{fig:linearity}$a$ shows the relation between the analog power and the digital variance normalized by the threshold as $(\sigma / V_0)^2$.
The best-fit linear regressions of $(\sigma/V_0)^2 = \alpha / V^2_0 W$ yielded $\alpha/V^2_0 = 2.548 \pm 0.001$ and $\alpha/V^2_0 = 2.524 \pm 0.0007$ for bias offsets of $\mu/V_0 = 0$ and $0.13$.
Departure from the linear regression was 4.0\% and 4.6\% at the minimum and maximum input power of 0.03 and 7.9 mW in the case of zero bias offset.
The departure in the biased case is not significantly different from the unbiased case.
The departures remain $<1.0$\% in terms of peak-to-peak in the input power range of 0.09 and 1.8 mW, that yields a dynamic range of 13 dB. 

\subsection{Bias endurance}
An additional linearity test was taken under a bias voltage in the input analog signal.
Instead of adding a bias voltage in the input analog signal, we adjusted the threshold voltages of the ADC.
The relative bias shift, $\mu / V_0$, were estimated by the statistics of digitized signals.

\begin{figure}[h!]
\begin{center}
\FigureFile(80mm,50mm){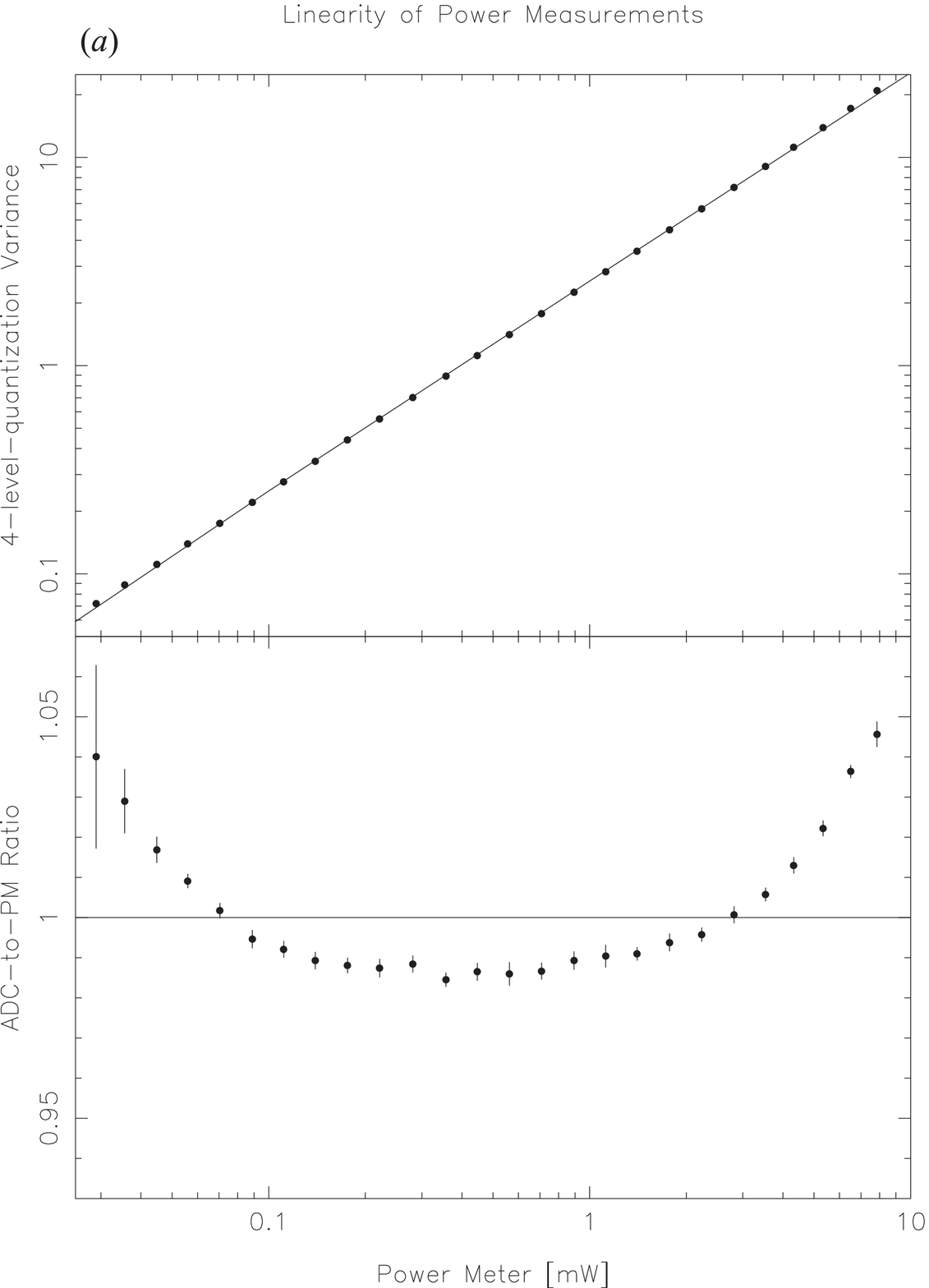}
\FigureFile(80mm,50mm){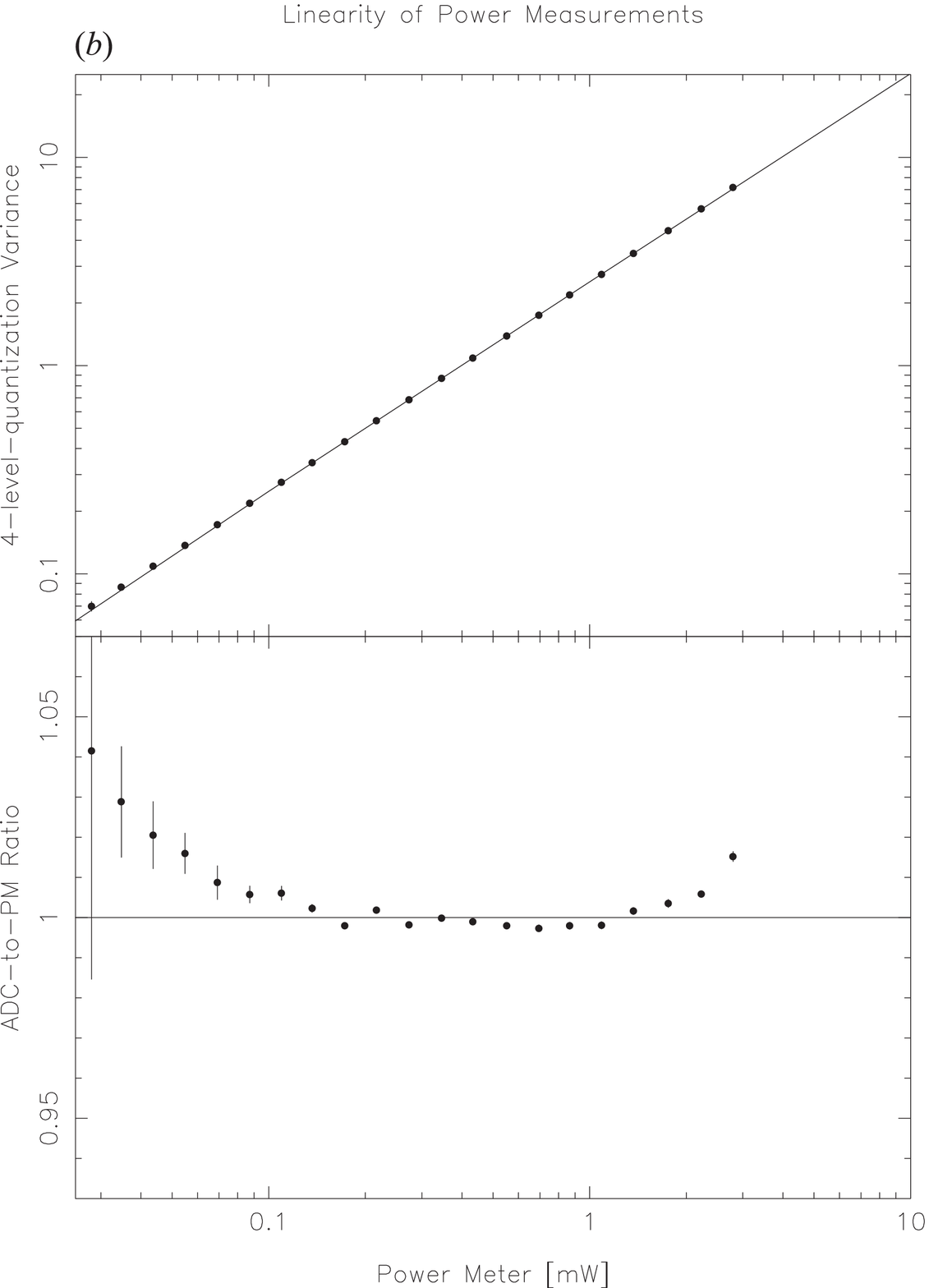}
\caption{($a$) Linearity without offset bias. ({\it Upper Panel}): Relation between the analog input power, $W$ [mW], and the variance normalized by the threshold voltage, $(\sigma / V_0)^2$, calculated from the statistics of digitized signals. The solid line indicates the best-fit linear regression of $(\sigma / V_0)^2 = \alpha W / V^2_0$, where $\alpha/V^2_0 = 2.548 \pm 0.0001$ mW$^{-1}$.
({\it Bottom}): Ratio of the variance to the linear regression.
The vertical axis indicates $\sigma^2 / (\alpha W)$.
The departure from the linear regression becomes larger, up to 4.0\% and 4.6\% at the lowest and the highest input power, respectively.
It remains almost constant with the peak-to-peak range of 1.0\% in the input power range between 0.09 and 1.8 mW, that yields a dynamic range of 13 dB.
($b$) Linearity with a bias offset of $\mu/V_0 = 0.13$. The linear regression shows $\alpha = 2.524 \pm 0.0007$. The departure from the linear regression remains $<1.0$ \% (p-p) in the input power range between 0.09 and 2.2 mW that yields a dynamic range of 14 dB. 
}
\label{fig:linearity}
\end{center}
\end{figure}

Figure \ref{fig:linearity}$b$ shows the linearity relation, as same as figure \ref{fig:linearity}$a$, with the bias shift of $\mu / V_0 = 0.13$.
In this test the input power was changed from $-40$ through $-20$ dBm.
The best-fit linear regression was $\alpha W / V^2_0 = (1 \pm 0.0007) (\sigma / V_0)^2$.

\begin{figure}[h]
\begin{center}
\FigureFile(80mm,50mm){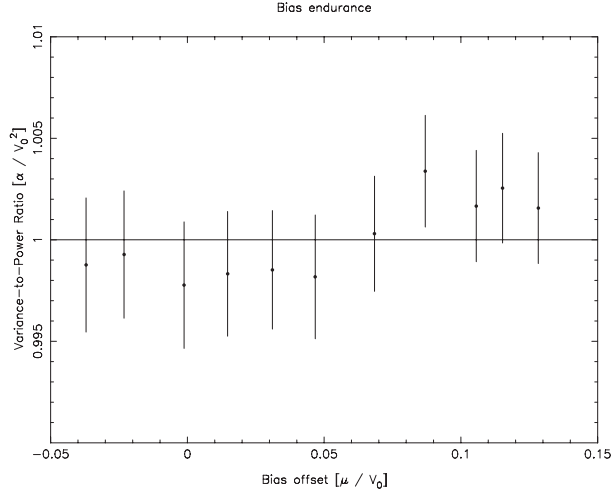}
\caption{Bias endurance of power measurements. The variance-to-power ratio $(\sigma/V_0)^2/\alpha W$ was measured at several bias offset between $-0.04 \leq \mu / V_0 \leq 0.128$. Throughout bias offset, the ratio was almost unity with no errors beyond $0.4$\%.}
\end{center}
\label{fig:bias_robustness}
\end{figure}

The ratios, $\alpha$, in various bias voltages are plotted in figure \ref{fig:bias_robustness}.
No significant departure greater than 0.4\% from 1 was found in the tested relative bias range of $-0.04 \leq \mu/V_0 \leq 0.13$.

\subsection{Analog and digital $T_{\rm sys}$ measurements}
We examined accuracy of digitally-estimated $\hat{T}_{\rm sys}$ by comparing with $T_{\rm sys}$ of the conventional method, in various elevations.
$W_{\rm hot}$ was measured once at the elevation of $88^{\circ}$, that was the upper limit of the 6-m telescope.
$W_{\rm sky}$ was taken at $88^{\circ}$ and from $85^{\circ}$ through $10^{\circ}$ by a $5^{\circ}$ step.
Statistics of digital samples was also acquired at each elevation.

\begin{figure}[h!]
 \begin{center}
   \FigureFile(80mm,80mm){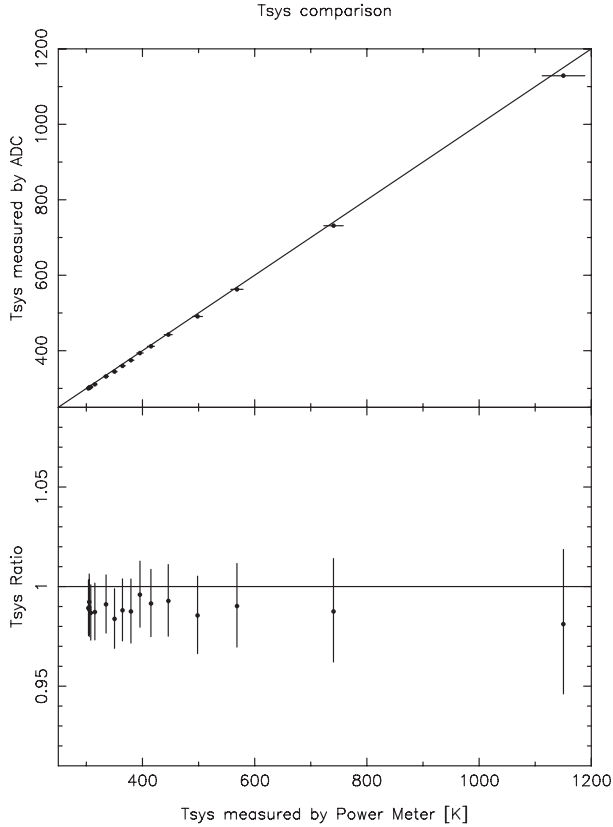}
 \end{center}
 \caption{({\it Top:}) Comparison of digitally measured $\hat{T}_{\rm sys}$ with the conventional $T_{\rm sys}$ using a power meter. $T_{\rm sys}$ ranged between 303.7 K and 1150.4 K for elevation angles of $88^{\circ}$ and $5^{\circ}$ of the Kagoshima 6-m radio telescope. ({\it Bottom:}) The ratio of digital $\hat{T}_{\rm sys}$ to conventional $T_{\rm sys}$ measurements. Relative error from unity was less than $1.9$\%. }
\label{fig:tsys_comparison}
\end{figure}

Figure \ref{fig:tsys_comparison} shows the comparison between $T_{\rm sys}$ and $\hat{T}_{\rm sys}$.
The linear regression resulted in $\hat{T}_{\rm sys} = (0.988 \pm 0.019) T_{\rm sys} + 0.3 \pm 6.6$.
Departure of $\hat{T}_{\rm sys} / T_{\rm sys}$ from unity was less than 1.9\% in whole $T_{\rm sys}$ range from $303.7$ through $1150.4$ K.

\section{Discussion}

\subsection{Performance in system noise measurements}
As shown in figure \ref{fig:linearity} $\sigma^2$ proportionally responses to the input power. Departure from the ideal linear response is $4$\% at the lowest input power and $4.6$\% at the maximum power of 7.85 mW (+9 dBm).
The departure from the diagonal line is within 1\% (between $-0.55$ and $-1.55$ \%) if we use the input power within 0.09 mW to 1.8 mW. Thus, the linearity is as good as 99\% in this dynamic range of 13 dB without any correction.
This error range is comparable with the accuracy of the power meter, specified as $\pm 0.5$\%.

The departure increases as input power exceeds 1.8 mW. This behavior can be caused by saturation in the power meter. Relevant calibration curve would reduce the departure and overcome the nonlinearity.

Since the power ratio between {\it hot} and {\it sky} inputs is usually less than 10 dB, the linearity without correction is enough for Tsys measurements with the accuracy of 1\%.

The dynamic range of 13 dB for the 99\% linearity is acceptable for usual Tsys measurements using hot and cold loads.
For lower-noise system, wider dynamic range is needed.
When Tsys is better than 32 K, hot-sky ratio exceeds 10 dB.
Relevant calibration curve would expand the dynamic range and allows us to measure such low-noise system. 

\subsection{Accuracy in system noise measurements}
Error in ${T}_{\rm sys}$ is estimated from equation \ref{eqn:tsys_power} as
\begin{eqnarray}
\left( \frac{\delta T_{\rm sys}}{T_{\rm sys}} \right)^2 = \left(\frac{W_{\rm sky}}{W_{\rm hot}-W_{\rm sky}}\right)^2 \left[  \left( \frac{\delta W_{\rm sky}}{W_{\rm sky}} \right)^2 + \left( \frac{\delta W_{\rm hot}}{W_{\rm hot}} \right)^2 \right], \label{eqn:tsys_error}
\end{eqnarray}
where $\delta W_{\rm sky}$ and $\delta W_{\rm hot}$ are errors of power in {\it sky} and {\it hot} circumstances.
In the case of digital $\hat{T}_{\mathrm{sys}}$ measurements, the error in power consists of error in $\sigma^2$ and variance in $\alpha$ within certain range of {\it hot} and {\it sky} powers.

Statistical ambiguity in bit distribution causes the error in $\sigma^2$.
Uncertainty in $n_{\tt ml}$ is estimated by variance of the binominal distribution, $n P_{\tt ml} (1 - P_{\tt ml})$, so that the error in $P_{\tt ml}$ is given by
\begin{eqnarray}
\sigma^2_{P_{\tt ml}} = \frac{P_{\tt ml} (1-P_{\tt ml})}{n}. \label{eqn:probability_error}
\end{eqnarray}
The number of sample, $n$, equals to the product of the sampling rate and the integration time.
In our experiment the sampling rate of 512 Msps and the integration of 1 second brought $n = 5.12\times 10^8$ and resulted in $\sigma^2_{P_{\tt ml}} \sim 10^{-9}$.

Variance in $\alpha$ is estimated by the results of the linearity test as shown in figure \ref{fig:linearity}, that is less than 1\% within the range of 10 dB.
This term is much larger than the error in $\sigma^2$, hence, is the major determinant in $\delta W_{\rm sky}$ and $\delta W_{\rm hot}$ unless $P_{\tt ml} < 10^{-5}$.
Using equation \ref{eqn:tsys_error} the total error in digital $T_{\rm sys}$ measurement is estimated to be $\sim 1$\%.
As shown in figure \ref{fig:tsys_comparison}, $\hat{T}_{\rm sys}$ measure by the bit distribution coincides the conventional measurements within 1.88\%.
The maximum error appears at the lowest elevation of $10^{\circ}$ where the factor $W_{\rm sky} / (W_{\rm hot}-W_{\rm sky})$ is largest.
The magnitude of departure is reasonable as estimated in equation \ref{eqn:tsys_error}, with some additional error caused by the power meter and different signal path after the power splitter.

\subsection{Robustness against bias}
The power ratio differs only 0.2\% when added bias of $-0.037$ to $+0.128$ $V_0$.
The results simply show that new digital power measurements is robust enough against input voltage bias.
As shown in equation \ref{eqn:probability_equations} $\sigma$ and $\mu$ can be well orthogonally determined by the non-linear least squares.
This is why $\hat{T}_{\rm sys}$ was not significantly affected by the bias.
Larger bias, $|\mu| \gg  |V_0|$, however, can cause very small probability in $P_{\tt 00}$ or $P_{\tt 11}$ that can result in relatively large error in $\hat{T}_{\rm sys}$.

\subsection{Inconvenience as a power meter}
Although the new method works for the $T_{\rm sys}$ measurement, it is not suitable as a general-purpose power meter.
The input signal voltage is presumed to follow the normal distribution in the estimation of $\sigma$.
The assumption is improper unless the input signal is a white noise, such as the system noise.
A monochromatic non-biased signal at a particular frequency with the amplitude of $V_0$, for instance, gives a probability density function as $p(V) = \frac{1}{\pi} (V^2_0 - V^2)^{1/2}$, that is different from the normal distribution.
A 4-level ADC with the threshold voltage of $v_0$ will exhibits a non-linear response to such a signal and the bit distribution obviously freezes when $V_0 \leq v_0$.
The bit distribution is effective only when the probability distribution is known and spans the range of digital states.

\section{Summary}
We examined performance of $T_{\rm sys}$ measurement using a 2-bit ADC, and verified that the new method is available as an alternative of conventional measurement using a power meter.
The statistics of digitized samples showed a sufficiently linear response whose departure from power-meter measurement was less than 1\% in a dynamic range of 13 dB.
Although we added bias offsets in input voltage up to 0.128 $V_0$, the linear response was kept in the same dynamic range.
The linearity and dynamic range yielded accurate $T_{\rm sys}$ measurements that differed $< 1.9$\% from conventional measurements using a power meter at various antenna elevation angles.

Modern radio telescopes and interferometers that are equipped with a high-speed ADC with 2-bit quantization can use this new method.
ADCs with 3- or more-bit quantization are expected to perform better accuracy and a wider dynamic range with an appropriate calculation of bias offset and standard deviation similar to this method.
This method is available without any additional hardware.
Since this method counts digitized signals that will be used in spectroscopy or interferometry, it would be less affected by systematic errors that is cased by non-linearity and different passband characteristics in a different signal path to a power meter used in conventional $T_{\rm sys}$ measurements.
Thus, this method can be a new standard of $T_{\rm sys}$ measurements in radio astronomy.

We thank Dr. J. D. Romney who reviewed this paper and gave positive suggestions to improve the manuscript. 
The Kagoshima 6-m telescope is supported by the National Astronomical Observatory of Japan and operated by staffs and students of the Kagoshima University. 


%

%

\end{document}